
\def\epskip{\vskip-5pt\noindent}

%

\input gr92.sty

\rightline{\bf TO--JLL--P 2/92}
\rightline{Turin, December 1992}
\vskip1truecm

\title{A Model of Topological Affine Gravity
\goodbreak in Two Dimensions}
\short{A Model of Topological Affine Gravity$\ldots$}

\authors{\first{Marco FERRARIS}{1}\etal{Mauro FRANCAVIGLIA}{2}
\last{Igor VOLOVICH}{2,\flat}}

\addresses{\addr{1}
{Dipartimento di Matematica, Universit\`a di Cagliari, \epskip
Via Ospedale 72
, 09134 CAGLIARI (ITALY)}
\addr{2}
{Istituto di Fisica Matematica ``J.--L. Lagrange'', Universit\`a di Torino,
\epskip Via C. Alberto 10, 10123 TORINO (ITALY)}}
\support{\flat}{Permanent address: Steklov Mathematical Institute,
Russian
Academy \epskip of Sciences, Vavilov St. 42, GSP--1, 117966 MOSCOW (RUSSIA)}

\summary{
A model of two--dimensional gravity with an action depending only on a
linear connection is considered. This model is a topological one, in
the sense that the classical action  does not cont
ain a metric or zweibein
at all.
A metric and an additional vector field are instead introduced in the process
of solving equations of motion for the connection. They satisfy the constant
curvature
equation. It is shown that the general solution of these equations of motion
can be described by using the space of orbits under the action
of the
Weyl group in the functional space containing all pairs formed by a metric
and a  vectorfield.
It is shown also that this model admits an equivalent description by u
sing a family of
actions depending on the metric and the connection as independent variables.}
\vfill\eject
\magnification=1200

\section{1}{Introduction}

Two dimensional gravity provides a conceptual laboratory for
studying gravity in higher dimensions and it is a basis for
string theory. As is well known there are no
Einstein equations in two dimensions. One considers instead the equation of
constant curvature

$$R=\Lambda \eqno (1.1)$$

\noindent where $R$ is the scalar curvature of a metric $g$,
$R=R(g)$ and
$\Lambda$ is a constant.

To derive eq. (1) from a variational principle Teitelboim [1] and
Jackiw [2] suggested the action

$$S(g,\Phi)=\int  (R-\Lambda)\Phi \sqrt{g}\ d^2x \eqno (1.2)$$

\noindent
containing an additional scalar field $\Phi$. Unfortunately, the field $\Phi$
has no direct geometrical origin within a (two--dimensional)
space--time. This model was further
considered in [3,4]. A gauge theoretical formulation of the model was
given in [5,6,7] by using an action depending on a zw
eibein $e$, a
spin connection $\omega$ and the scalar field $\Phi$ itself.

A different approach to two--dimensional gravity was suggested
in [8], based on the action

$$
S(e,\omega)=\int (\gamma R^2 +\beta T^2 +\lambda)\ det(e)\ d^2x \eqno
(1.3)
$$

\noindent
where $T$ is the torsion of a spin connection $\omega$, $R$ denotes the scalar
curvature constructed using the
Riemann
tensor of $\omega$ and a zweiben $e$, while $\gamma, \beta\  {\rm and\ }
\lambda$ are constants. Here $\omega$ and $e$ are the onl
y dynamical variables
and they have a geometrical origin.
The model was investigated in [9--15]. It was shown to contain
eq. (1) as a sector in the space of solutions.

In this paper
we shall propose and consider a new model, whose classical action
is ``purely affine'', i.e. it
does not contain a metric or zweibein at all, but only
a linear connection $\Gamma$ as a
basic field. In this sense
it is therefore a ``purely topological'' theory.  This model is rather
different from models of topological fiel
d theories which have been
investigated
in recent years; for a review see e.g. [16]. Purely affine theories
in four dimensions were discussed by Einstein [17],
Eddington [18]
and Schr\"odinger [19]; for more recent developments see e.g. [20]
and references quoted therein.

It seems to us that such an
approach could be suitable for investigating the still mysterious
Witten's ``unbrocken phase'' [21] with $<g_{\mu \nu}>=0$. We will
see in fact how a metric can be constructed out of the basic field $\Gamma$

Actually, the symmetric part of the Ricci tensor $G_{\mu\nu} (\Gamma)$
will play the role of the metric $g_{\mu\nu}=G_{\mu\nu}
(\Gamma)$, as
is usual in purely affine theories (see [20]). In this way one can
consider the metric $g$ as a "bound state" or
"condensate"
of the more fundamental field $\Gamma$. In other words, one treats
here the graviton as a composite particle.

A metric and an additional vectorfield are introduced in the process
of solution of the equations of motion for the connection ent
ering the
action. They satisfy a constant
curvature equation.
We will show that the general solution of these equations of
motion for the
connection can be in fact described by using the space of orbits under
the action
of the
Weyl group in the functional space of all pairs formed by a metric and a
vector field.
It will also
be shown that this model admits an equivalent metric--affine description,
i.e. a description in terms of a (family of)
actions depending on a metric and a connection as independent var
iables.

\section{2}{The Action and the Equations of Motion}

Let us consider the following action

$$
S(\Gamma)=k\int_M\sqrt{|det( G_{\mu\nu}(\Gamma))|}\ d^2x   \eqno (2.1)
$$

\noindent where $k$ is a constant and $M$ is a two--dimensional manifold
endowed with a symmetric linear connection $\Gamma^{\sigma}_{\mu\nu},
\mu,\nu,\sigma =1,2$.  One defines the Riemann curvature tensor by

$$
R_{\mu \nu \sigma}^{\lambda}=\partial_{\nu} \Gamma_{\mu
\sigma}^{\lambda} - \partial_{\sigma}\Gamma_{\mu
 \nu}^{\lambd
a} + \Gamma_{\alpha\nu}^{\lambda} \Gamma_{\mu
\sigma}^{\alpha} - \Gamma_{\alpha \sigma}^{\lambda}
\Gamma_{\mu \nu}^{\alpha} \eqno (2.2)
$$

\noindent The Ricci tensor is defined by

$$
R_{\mu \sigma}(\Gamma) = R_{\mu \nu \sigma}^{\nu} \eqno (2.3)
$$

\noindent and $G_{\mu\nu}$ denotes its symmetrization

$$
G_{\mu\nu}(\Gamma)= R_{(\mu\nu)}(\Gamma)\eqno (2.4)
$$

We will also use the notation

$$
G=G(\Gamma)=|det( G_{\mu\nu}(\Gamma))|
$$

\noindent and define the covariant derivative $\nabla_\alpha$ with r
espect to
$\Gamma$ acting on a vectorfield $v^\mu$ as

$$\nabla_{\alpha}v^\mu = \partial_\alpha v^\mu +\Gamma^\mu
_{\alpha\sigma}v^\sigma
$$

Let us now derive the equations of motion as Euler--Lagrange equations of
the action (2.1). We assume
$det( G_{\mu\nu})\not=0$.  By using the standard formula for variations

$$
\delta G_{\mu\nu} =\nabla_\alpha (\delta \Gamma^\alpha_{\mu\nu}-
\delta^\alpha_{(\nu}\delta\Gamma^\sigma_{\mu)\sigma} )
$$

\noindent one gets from (2.1) the following equations of motion

$$
\nabla_\alpha (\sqrt{G} G^{\mu\nu}(\Gamma))=0  \eqno  (2.5)
$$

\noindent Here $ G^{\mu\nu}=  G^{\mu\nu}(\Gamma) $  is the inverse matrix of
$G_{\mu\nu}$:

$$
G^{\mu\nu}G_{\nu\sigma}=\delta_\sigma^\mu
$$

\section{3}{Solution of the Equations of Motion}

We have now the system of equations (2.5) for the linear connection $\Gamma$.
The general solution of these equations of motion
can be represented in the following form:

$$
\Gamma_{\mu\nu}^\sigma=\Gamma_{\mu\nu}^\sigma(g,B)=\gamma_{\mu\nu}^\sigma(g
)
+w_{\mu\nu}^\sigma(g,B)   \eqno(3.1)
$$

\noindent
where $g _{\mu\nu}$ and $ B_\mu $ are respectively a (non--degenerate)
symmetric tensor and a vectorfield
satisfying the equation

$$
R(g)-D_\sigma B^\sigma =2\Lambda \eqno (3.2)
$$

\noindent $\Lambda$ being a non-zero constant,

$$\gamma_{\mu \nu}^{\sigma}(g)=
{1\over2}g^{\sigma \alpha}(\partial_{\mu}g_{\nu \alpha}
+\partial_{\nu}g_{\mu \alpha}-\partial_{\alpha}g_{\mu \nu})
\eqno (3.3)
$$

\noindent is the Levi--Civita connection of $g$ and $w_{\mu \nu}
$ is given by

$$
w_{\mu \nu}^{\sigma}(g,B)={1\over 2}(B_\mu\delta_\nu
^\sigma+B_\nu\delta_\mu^\sigma -B^\sigma g_{\mu\nu})
\eqno (3.4)
$$

\noindent Here $D_\sigma$ is the covariant derivative
with respect to $\gamma (g)$,
$R(g)$ is the scalar curvature of $g$

$$
R(g)=R_{\mu\nu}(g)g^{\mu\nu}
$$

\noindent and
$R_{\mu\nu}(g)$ denotes the Ricci tensor of the connection $\gamma(g)$.

To prove our claim let us represent eq. (2.5) in the form

$$
\nabla_\mu G_{\alpha\beta}=(-\Gamma ^\sigma_{\mu\sigma}+
{1\over
 2}G^{\sigma\tau}\partial_\mu G_{\sigma\tau})G_{\alpha\beta}
\eqno (3.5)
$$

\noindent Setting then

$$
g_{\alpha\beta}={1\over\Lambda}G_{\alpha\beta}(\Gamma)  \eqno (3.6a)
$$

$$
B_\mu=\Gamma ^\sigma_{\mu\sigma}-{1\over 2}g^{\sigma\tau}\partial_\mu
 g_{\sigma\tau}   \eqno (3.6b)
$$

\noindent one can rewrite eq. (3.5) as

$$
(\nabla_\mu +B_\mu)g_{\alpha\beta}=0  \eqno (3.6c)
$$

\noindent
were $\Lambda$ is an arbitrary constant, $\Lambda\not=0.$ Clearly,
the system of eqs. (3.6) for $\Gamma,g,B$ is equi
valent to
eq. (3.5) or (2.5) for $\Gamma$.

Let us then consider the system (3.6). It is not hard to see that the
general solution of eq. (3.6c) for the connection
$\Gamma$ is

$$
\eqalign{\Gamma_{\mu \nu}^\sigma
&={1\over2}g^{\sigma \alpha}((\partial_{\mu}+B_\mu )g_{\nu \alpha}
+(\partial_{\nu}+B_\nu )g_{\mu \alpha}-(\partial_{\alpha}+B_\alpha )
g_{\mu \nu})\cr
 &=\gamma_{\mu \nu}^{\sigma}(g)+w_{\mu \nu}^{\sigma}(g,B) \cr} \eqno (3.7)
$$

\noindent Equation (3.6b) follows from (3.7).

One needs now to so
lve eq. (3.6a).
Using the definition (2.3) of the Ricci tensor for the connection
(3.7) one gets

$$
R_{\alpha\beta}=G_{\alpha\beta}+F_{\alpha\beta}
$$

\noindent where

$$
G_{\alpha\beta}={1\over2}(R(g)-D_\sigma B^\sigma)g_{\alpha\beta}
 \eqno (3.8)
$$

\noindent and

$$
F_{\alpha\beta}={1\over2}(\partial_\alpha B_\beta - \partial_\beta B_\alpha)
$$

Comparing eqs. (3.8) and (3.6a) we see that $g$ and $\Gamma$
satisfy eq.
(3.6a) if the metric $g$ and the vectorfield $B$ satisfy the equation

$$
R(g)-
D_\sigma B^\sigma =2\Lambda
$$

\noindent i.e., eq. (3.2).

Thus, we have
shown that the general solution of eq. (2.5) can be represented in the
form (3.7),
where $
g_{\alpha\beta}$ and $ B_\mu$ satisfy eq. (3.2) \hskip5truecm (Q.E.D).
\vskip1truecm

Notice that if one introduces the curvature of $\Gamma$

$$
{\cal R}=R(\Gamma,g)=g^{\mu\nu}G_{\mu\nu}(\Gamma)
$$

\noindent which, using eq. (3.1), is equal to

$$
{\cal R}={\cal R}(g,B)=
R(\Gamma(g,B),g))=R(g)-D_\sigma B^\sigma=2\Lambda   \eqno (3.9)
$$

\noin
dent one can say that the action (2.1) leads to the constant
curvature equation (3.9).

Let us remark
that from (3.7) it follows that the pair $(g_{\alpha\beta},B_\mu)$
and any of its Weyl transformations
$(e^\lambda g_{\alpha\beta},B_\mu-\partial_\mu\lambda)$,
where $\lambda$ is an arbitrary scalar
function,
give the same
linear
connection $\Gamma$. This means that one can parametrize the space
of solutions
of eq. (2.5) for $\Gamma$ by using the space of conformal orbits
of pairs $(g,B)$, i.e. the quoti
ent of the space of all pairs $(g,B)$ under
the action of the Weyl group.

Let us also
remark that eq. (3.2) is less restrictive than eq. (1.1). In (3.2) one
has in fact one equation
for five
functions $(g_{\alpha\beta}, B_\mu)$, whereas in (1.1) one has one
equation
for only three functions  $(g_{\alpha\beta})$. When considering
equations (3.1)--(3.2)
one should also take into account the Weyl symmetry.

\section{4}{Metric--Connection Formalism}

We show here that the model discussed can be also be descri
bed by using
a metric--affine action depending on a metric and a connection
as independent variables.
Let us consider the following action

$$
S(\Gamma,g)=\int_M L(R)\sqrt{g}\ d^2x \eqno (4.1)
$$

\noindent where $M$ is a two--dimensional manifold endowed with a metric
$g_{\mu\nu}$ and a symmetric linear connection $\Gamma_{\mu\nu}^\sigma$.
Such an action on n--dimensional manifolds, $n\geq 3$, was considered
in [22]. We shall use notaion (2.2) for the Riemann tensor, (2.3) for the
Ricci tensor, (2.4) for
its symmetric part and the following

$$
R = R(\Gamma,g)=R_{\mu\sigma}(\Gamma)g^{\mu\sigma}  \eqno (4.2)
$$

Here $L$ is a given function of one real variable, which we assume to
be analytic on the real line.

The Euler--Lagrange equations for the action (4.1) with respect to
$ g$ and $ \Gamma$
can be written in the following form

$$
L^{\prime} (R)R_{(\mu \nu )}(\Gamma)-{1\over2}L(R)g_{\mu
\nu}=0\eqno (4.3)
$$

$$
\nabla_{\alpha}(L^{\prime} (R)\sqrt{g} g^{\mu\nu})=0\eqno (4.4)
$$

\noindent where $ \nab
la_{\alpha}$ is the covariant derivative with respect
to $ \Gamma$.

Multiplying eq. (4.3) by $ g^{\mu\nu}$ one obtains

$$
L^{\prime} (R)R-L(R)=0 \eqno (4.5)
$$

\noindent If eq. (4.5) is identically satisfied we have

$$
L(R)=cR\eqno (4.6)
$$

\noindent where $c$ is an arbitrary constant.

In all other cases, for a given analytic function $ L(R)$ eq. (4.5)
can have no more than a countable set of solutions $R=c_i, i=1,2,...$
where $c_i$ are
constants. Let us further assume that $L^{\prime}(c_i)\not=0$ fo
r any
$i=1,2,...$ Consider then a solution

$$
R=c_i \eqno (4.7)
$$

\noindent so that eq. (4.4) takes the form

$$
\nabla_{\alpha}(\sqrt{g} g^{\mu \nu})=0 \eqno (4.8)
$$

This last equation (4.8) is in fact equivalent to the pair of equations
(3.6b) and (3.6c), so that one can again apply the considerations
of section 3. Hence, one gets the general solution of eq. (4.8) under
the form (3.7), with the metric $g_{\alpha\beta}$ and the vectorfield
$B_\mu$ satisfying the equation

$$
R(g)-D_\sigma B^\sigma=2\L
ambda   \eqno (4.9)
$$

\noindent where $\Lambda$ now is equal to

$$
\Lambda =L(c_i)/2L^{\prime}(c_i)=c_i/2  \eqno (4.10)
$$

This shows ``universality'' of the constant curvature equation (4.9),
in the sense that any analytic Lagrangian $L(R)$, with
$R=R(\Gamma,g)$, subject to the only restrictions $L^{\prime}(c_i)\not=0$
and $L(R)\not=cR$,
will generate under independent variations of $\Gamma$ and $g$ a
system of fields $(g_{\alpha\beta},B_\mu)$ satisfying equation
(4.9). Eq. (4.9) should be compared wit
h the expression
(3.9) for purely--affine theories.

As an example, let us consider the quadratic Lagrangian

$$
L(R)=aR^2 + bR + \lambda \eqno (4.11)
$$

where $a,b$ and $\lambda$ are constants ($a\not=0$).
In this case one has the eq. (4.9) with

$$
\Lambda=\pm{1\over2}\sqrt{{\lambda \over a}} \eqno (4.12)
$$

It is interesting to notice that the exceptional Lagrangian
$L(R)=cR(\Gamma,g)$ gives here non trivial restrictions to
the set of dynamical fields, because in this case from eq. (4.4) it
follows
that the linear connection $\Gamma$ should have the form (3.7).
Therefore, the
general solution of equations of motion will depend on five independent
functions
$(g_{\alpha\beta},B_\mu)$ instead of the nine original functions
$(g_{\alpha\beta},\Gamma^\sigma_{\mu\nu})$. We also recall that,
in contrast, in the purely--metric formalism the Lagrangian $L(R)=cR(g)$
does not give any restriction at all onto dynamics.

\section{5}{Conclusions}

The action (2.1) together with its equations of motion (2.5) for a

linear connection $\Gamma$ was investigated in this paper. It was
shown that the general solution of these equations can be
represented in the form (3.1), with fields $g$ and $B$ satisfying
equations (3.2); where $\Lambda$ is an arbitrary (non-zero) constant.
The metric $g_{\mu\nu}$ and the vectorfield $B_\mu$ appear in the
process of solutions of equations for $\Gamma$.

One has a sort of ``dynamical duality''. One can in fact describe the
same system either in terms of the linear connection
$\Gamma$ usin
g the purely--affine action (2.1), or alternatively by using
the pair $(g,\Gamma)$ with the metric--affine action (4.1)
under the assumption $L^{\prime}(c_i)\not=0$. The duality is
explicitly described by eqs. (3.1), which gives $\Gamma$ in terms
of $g$ and $B$, together withe inverse relations (3.6a) and (3.6b),
which can be in fact rewritten as follows:

$$
\eqalign{
g_{\alpha\beta}=
&g_{\alpha\beta}(\Gamma)={1\over\Lambda}G_{\alpha\beta}(\Gamma)  \cr
B_\mu=&B_\mu(\Gamma)=\Gamma ^\sigma_{\mu\sigma}-{1\over
 2}G^{\sigma\tau}\partial_\mu
 G_{\sigma\tau} \cr
} \eqno (5.1)
$$

\noindent
where $G_{\sigma\tau}=G_{\sigma\tau}(\Gamma)$ is given by eq. (2.4).

It should be stressed that to any given affine action (2.1)
with a fixed constant $k$ there corresponds a whole family
of metric--affine actions (4.1), because the constant
$\Lambda$ appearing in (3.2) is arbitrary. For example, one
can consider the family of quadratic Lagrangians (4.11)
with a variable parameter $\lambda$ and $a\not=0$ fixed, to obtain
$\Lambda
(\lambda)$ as in eq. (4.12).

Many natural questions there appear. For example: which is
the most natural action depending on the variables
$(g_{\alpha\beta},B_\mu)$ and leading to equations (3.2)?
An obvious generalization of Teitelboim--Jackiw action would
consist in the following

$$
S(g,B,\Phi)
=\int_M (R(g)-
D_\sigma B^\sigma -2\Lambda)\Phi \sqrt{g} \ d^2x \eqno (5.2)
$$

\noindent
which gives equations (3.2) under variations with respect to $\Phi$;
however, we did not yet investigate in detail the furt
her equations
ensuing from this action by taking variations with respect to $g$ and
$B$.

Notice moreover that the action (1.3) depends in fact on a vectorfield,
because in two dimensions one can always parametrize the spin connection
$\omega$ as follows

$$
\omega_\mu^{ab}=\varepsilon^{ab}B_\mu  \eqno (5.3)
$$

\noindent where $\varepsilon^{ab}$ is the totally skew--symmetric tensor
of Levi--Civita
and $B_\mu$ is an arbitrary vectorfield. It would be therefore interesting
to exploit the relations between t
he theories considered in [8--15],
depending on the action
(1.3), and the new topological model considered above. We
recall that a relation
between
the theory defined by the action (1.3) and standard topological field
theory was discussed in [15].

The quantum aspects
of the model considered in this paper, in view of anomalies ensuing from
Weyl's invariance and the symmetry breaking involved in the generation
of new fields $(g,B)$ out of the given connection $\Gamma$, which may
produce non--equivalence at t
he quantum level between affine and metric--affine
equations, should also
be investigated. We hope to address these problems in forthcoming papers.

\noindent{\bf Acknowledgements}
Thanks are due to G. Magnano for his useful comments. One of us (I.V.)
acknowledges the
hospitality of the Institute ``J.--L. Lagrange'' of Mathematical Physics
of the University of Torino and the support of G.N.F.M. of Italian C.N.R.
This work is
sponsored by G.N.F.M., M.U.R.S.T. (40\% \ Proj. ``Metodi Geometrici
e Probabilis
tici in Fisica Matematica''); one of us (M. Ferraris) acknowledges
also support from I.N.F.N.
\vskip2truecm
\references

\paper{1a}{C.~Teitelboim}
{Phys. Lett.}{B216}{1983}{41}

\book{1b}{C.~Teitelboim}
{in: Quantum Theory of Gravity}{ed. S.M.~Christensen}{Adam Hilger,1984}

\book{2a}{R.~Jackiw}
{in: Quantum Theory of Gravity}{ed. S.M.~Christensen}{Adam Hilger,1984}

\paper{2b}{R.~Jackiw}
{Nucl. Phys.}{B252}{1985}{343}

\paper{3}{M.~Henneaux}
{Phys. Rev. Lett.}{54}{1985}{959}

\paper{4}{M.~Abe and N.~Naka
nishi}
{Int. J. Mod. Phys.}{A6}{1991}{3955}

\paper{5}{T.~Fukuyama and K~.Kamimura}
{Phys. Lett.}{B160}{1985}{259}

\paper{6a}{A.H.~Chamseddin and D.~Wyler}
{Phys. Lett.}
{B228}
{1989}
{75}

\paper{6b}{A.H.~Chamseddin and D.~Wyler} {Nucl. Phys.}
{B340}
{1990}
{595}

\paper{7}
{K.~Isler and C.A.~Trugenberger}
{Phys. Rev. Lett.}
{63}
{1989}
{834}

\paper{8a}
{M.O.~Katanaev and I.V.~Volovich}{Phys. Lett.}
{B175}
{1986}
{413}

\paper{8b}{M.O.~Katanaev and I.V.~Volovich}{Ann. of Phys.}
{197}
{1990}
{1}

\pape
r{9a}
{M.O.~Katanaev}
{J. Math. Phys.}
{31}
{1990}
{882}
\paper{9b}
{M.O.~Katanaev}
{J. Math. Phys.}
{32}
{1991}
{2483}

\paper{10}
{K.G.~Akdeniz, A.~Kizilersu and E.~Rizaoglu}
{Phys. Lett.}
{B215}
{1988}
{81}

\paper{11}
{K.G.~Akdeniz, O.F.~Dayi and A.~Kizilersu}
{Mod. Phys. Lett.}
{A7}
{1992}
{1757}

\paper{12a}
{W.~Kummer and D.J.~Schwarz}
{Phys. Rev.}
{D45}
{1992}
{3628}

\paper{12b}
{W.~Kummer and D.J.~Schwarz}
{preprint TUW--92--03}
{}
{1992}
{}

\paper{13}
{H.~Grosse, W.~Kummer, P.~Presneider and D.
J.~Schwarz}
{preprint TUW--92--04}{ }{1992}
{}

\paper{14}
{T.~Strobl}
{preprint TUW--92--07}{}
{1992}{}

\paper{15}
{N.~Ikeda and K.--I.~Izawa}
{preprint RIMS--888}{}
{1992} {}

\paper{16}
{D.~Birmingham, M.~Blau, M.~Rakowski and G.~Thompson}
{Phys. Rep.}
{209}
{1991}
{129}

\paper{17}
{A.~Einstein}
{Sitzungber. Preuss. Akad. Wiss.}{137}
{1923}
{}

\book{18}
{A.~Eddington}
{The Mathematical Theory of Relativity}
{}
{Cambridge Univ. Press, 1923}

\paper{19}
{E.~Schr\"odinger}
{Proc. R. Irish. Acad.}
{51A}
{
1947}
{163,205}

\paper{20a}
{M.~Ferraris and J.~Kijowski}
{Rend. Sem. Mat. Univ. Pol. Torino}
{41}
{1983}
{169}

\paper{20b}
{M.~Ferraris and J.~Kijowski}
{Lett. Math. Phys.}
{5}
{1981}
{127}

\paper{20c}
{M.~Ferraris, M.~Francaviglia and G.~Magnano}
{J. Math. Phys.}
{31(2)}
{1990}
{378}

\paper{21}
{E.~Witten}
{Nucl. Phys.}
{B311}
{1988-1989}
{46}

\paper{22}
{M.~Ferraris, M.~Francaviglia and I.~Volovich}
{preprint TO--JLL--P 1/92}
{}
{December 1992}
{}

\bye